\documentclass[9pt,times]{IEEEtran}
\usepackage{amsmath}
\usepackage{amsfonts}
\usepackage{amssymb}
\usepackage{graphicx}
\usepackage{multirow}
\usepackage{verbatim}
\immediate\write18{texcount -tex -sum  \jobname.tex > \jobname.wordcount.tex}

\title{Coding Estimation based on Rate Distortion Control of H.264 Encoded Videos for Low Latency Applications}
\author{Amitesh Kumar Singam\\
        \small $^{1}$IEEE Signal Processing Society \\
        \small $^{2}$0000-0002-2532-0989 \\
}

\begin{document}

\maketitle

\begin{abstract}
In the field of video processing, advancements in video compression at various temporal and spatial resolutions which are needed in our research to quantify estimation of video quality whereabouts within spatial and temporal domain itself. It was necessary in our research to study the impacts of related video coding conditions upon perceptual quality due to issue raised by User Experience community regarding poor coding. But most of research studies are concerned with coding distortions affected by mostly due to poor coding which address high spatio-temporal resolutions. This paper presents overall 120 test scenarios for video sequences having low spatial and temporal spectral information were studied at \cite{Pashike830965}. Finally we concluded that even after achieving consistency within subjective scores, we got relevant data consistency after refining subjective scores, so it is not poor coding its due channel capacity which was traced out by rate distortion control.
\end{abstract}
\section{Introduction}
In H.264/AVC coding structure, video processing is conducted at block level where each video frame is divided into blocks of pixels. Spatial redundancy is exploited in video frame by coding original blocks which employs spatial prediction, transform, quantization and entropy. Temporal dependencies are exploited between successive frames, so that only few changes between them are to be encoded, it is done by motion estimation and compensation. The remaining spatial redundancies in video frame are exploited by coding the residual blocks. H.264 employs different sizes, shapes and higher resolution blocks for motion estimation. It uses integer based transform which approximates DCT used in MPEG-2 for transform. H.264 standards are defined by five types of slices/frames, they are I, P, B, SI, SP. SI and SP slices are also called as switching slices, which are used for transitions between two different streams, it is an uncommon feature.
\paragraph{}The compression efficiency is improved by using H.264/AVC compare to previously adopted standards. In H.264/AVC, spatial correlations are reduced by transformation, bitrate is controlled by quantization, temporal correlation is reduced by motion compensation prediction and statistical correlation is reduced by entropy encoding \cite{4375937}.
\paragraph{}In general, fast algorithms based on objective quality metrics are used to estimate the degradation of video quality during transmission and compression. The introduction of impairments leads to degradation of video quality, these impairments are called as visible flaws and these defects are decomposed into set of perceptual features known as artifacts. In \cite{4203067} authors has conducted four experiments in which they studied information of artifacts at different strengths and combinations by generation of artifacts artificially. They compered it with real compression impairment which contains similar visibility and annoyance. In the combination of Blocky-Blurry, blocky artifact is at high strength whereas blurry artifact is at low strength. The artifacts such as blockiness, noisiness, ringing and blurriness do not have major effect on fitting parameters but has major effect on original content. This paper concluded after final analysis of four papers that parameters such as visibility and annoyance are linearly related to each other and positively correlated.
\section{Survey Related Works based on Artifacts of H.264 Standards}
In general, video or image compression techniques like JPEG, MPEG, and H.264 uses block based DCT since it provides high energy compaction. When the video or image is highly compressed with DCT technique results in blocking artifact which have discontinuous effect between neighbor blocks. Therefore blocking artifacts will be more visible if the bit rate is lower and it results in strong degradation of reconstructed image.
 \newline There are three methods for elimination of artifacts, called pre, post and in-loop processing. Post processing has advantages among three approaches. H.264/AVC standards provides new methods like variable block size motion estimation , 4x4 Block-based integer transform, in loop filter and spatial intra prediction which leads to increase in compression performance..
\paragraph{}H.264 standard can enclose intra coded frames for random access. Unlike previously appeared predictive coded frames, coding noise pattern for intra-coded frames is different and inter frame prediction is also different for I-frames compared to P-frames. Since it is not applied for intra coded frames, this causes discontinuation in coding noise pattern and induces Intra flicker which leads to degradation of video quality at lower bitrates. In order to overcome this problem, Senda et al. \cite{4106879} proposed a flicker suppression method using Detented Quantization algorithm which minimizes discontinuation in coding noise patterns  between P and I frames. Results shows that proposed method minimizes flicker artifacts  more than 50$ \%$ using JM encoder based on H.264 standards.
\paragraph{}ZhaoLin Zhang et al. \cite{5437841} proposed a new algorithm based on HVS characteristics for fast response on estimating the blocking artifacts of H.264 encoded video sequences . This algorithm is implemented with consideration of temporal blocky distortion among two successive frames and it is also well adapted to videos with in loop de blocking filter. Proposed method doesn't require reference video sequence and it is computationally efficient. Results show that evaluation of proposed method for estimating blocking artifacts are validated by comparing it with standard quality metrics such as PSNR, SSIM, and VQM. This paper concludes that the performance of proposed model can be used as Objective Quality metric in real time scenarios.
\section{Overview of Video Compression Techniques}
In wireless networks, an uncompressed video needs huge amount of bandwidth and storage. End user cost is proportional to availability of bandwidth and data transmission capacity in network or channel. Therefore, data transmitted in the network is compressed with very effective and lossy compression algorithms. For live video streaming,the most common compression standards are H.263 standardized by International Telecommunication Union (ITU), MPEG-4 part 2 standardized by International Organization for Standardization, H.264 which is also known as Advanced video coding and  MPEG-4 part 10 standardized  by International Organization for Standardization (ISO)/International Electrotechnical Commission (IEC) and ITU.
\paragraph{}The initial phase in video generation is sampling in spatial, temporal and color domain. Spatial sampling refers to number of pixels in each frame, Temporal domain sampling refers to resolution in number of pictures per second and color sampling domain provides color space like Gray Scale and RGB.
\paragraph{}At present, Video coding algorithms are intended to support a combination of temporal and spatial prediction along with transform coding. Each frame is split into macro blocks, these macro blocks are paradigm in frames. Paradigm represents subset of macro blocks to decode independently. In video Compression we have three classes of frames,  B-frames, I-frames, P-frames. They together are called as group of pictures. Since frames are segmented into macro blocks, I-frame is an intra-coded frame which contains intra macro blocks, P-frame is a predicted frame, B-frame is bi-predicted frame which contains intra and predicted macro blocks. A sequence of video  which contains I-frames, P-frame and B-frames.
\paragraph{}The main reasons for video compression are limited network bandwidth for real time video transmission and limitations in storage capacity. The factors should be considered during compression are quality, compression rate, complexity and delay.Video compression usually utilizes two basic compression techniques, Inter and Intra frame compression. Inter frame compression is compression between the frames and it is designed to minimize temporal redundancy. Intra frame compression is compression within an individual frame, it is designed to minimize spatial redundancy. We employed H.264 standards for video generation, scaling and decoding. 
\section{DATA SCREENING}
In our past research work, shahid et.al\cite{Shahid834579} considered the recommendations given by ITU-R BT 500-12 specifications within lab setup of our experiments. Particularly, the method we followed was Single stimulus continuous quality evaluation(SSCQ), where a test video sequence is shown once without presence of any explicit reference, corresponds to the reality where users see only the processed version of videos This subjective experiment was conducted in a lab set-up designed in accordance with ITU standards. A flat LCD screen with non-glare surface treatment was used for displaying the video sequences. The used monitor had resolution 1440x900 with 5ms response time and its color temperature was set at 6500K in sRGB mode. Other hardware includes a desktop HP system having 3 GHz AMD processor and 4 GB RAM. A comfortable seating arrangement was made for the subjects at three to four times the high of the display screen. 
\newline A software tool developed at the department was used to automate the process of presenting the videos in the center of the screen. Videos were played in a random order for each subject with insertion of the standard intervals (10 sec.) in between for grading. Viewers were not given the privilege to repeat any video and software front end had no controls available for the subjects to alter the intended processes in anyway. The software automatically stored the results in an excel sheet. The grading scale used was 0-100 and the scores were mapped to the 1-5 scale afterwards for further use. 
 \begin{figure}[ht]
\centering
\includegraphics[width=0.4\textwidth]{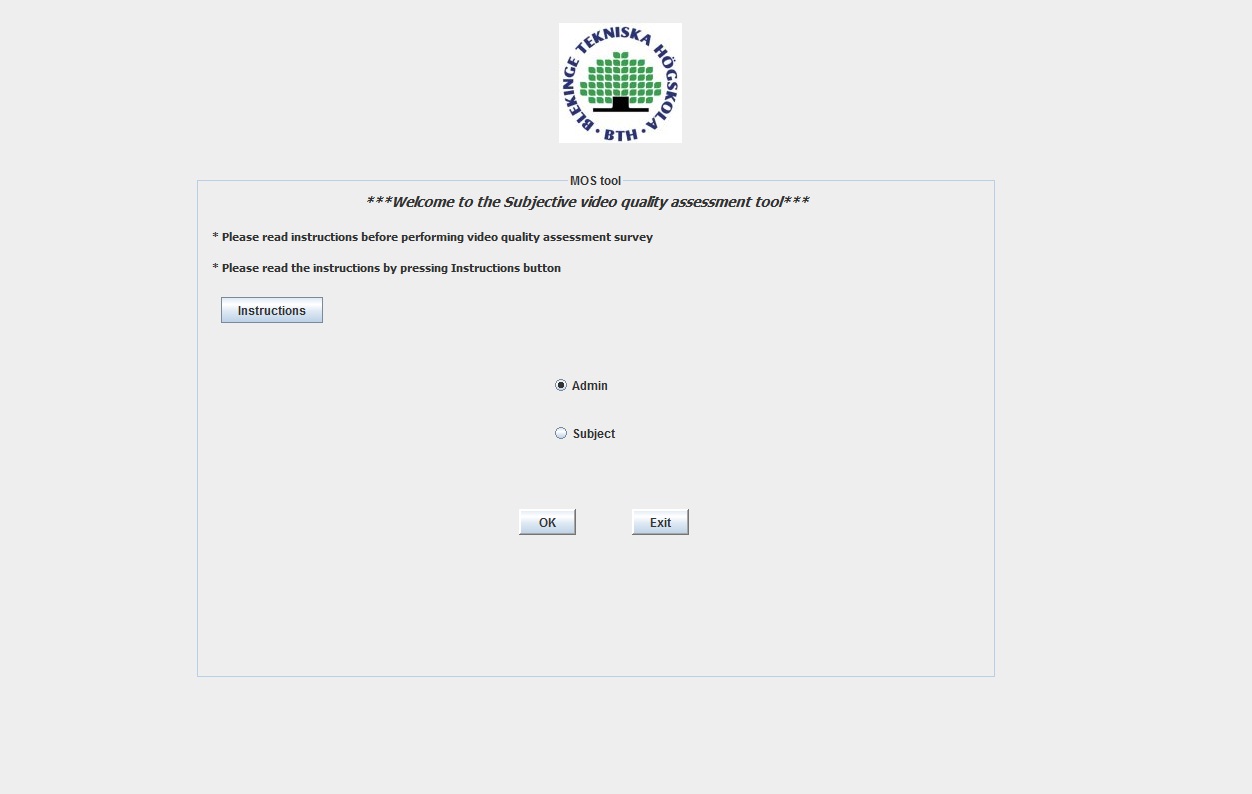}
\caption{Subjective analysis Tool developed by University}
\end{figure}
 \begin{figure}[ht]
\centering
\includegraphics[width=0.4\textwidth]{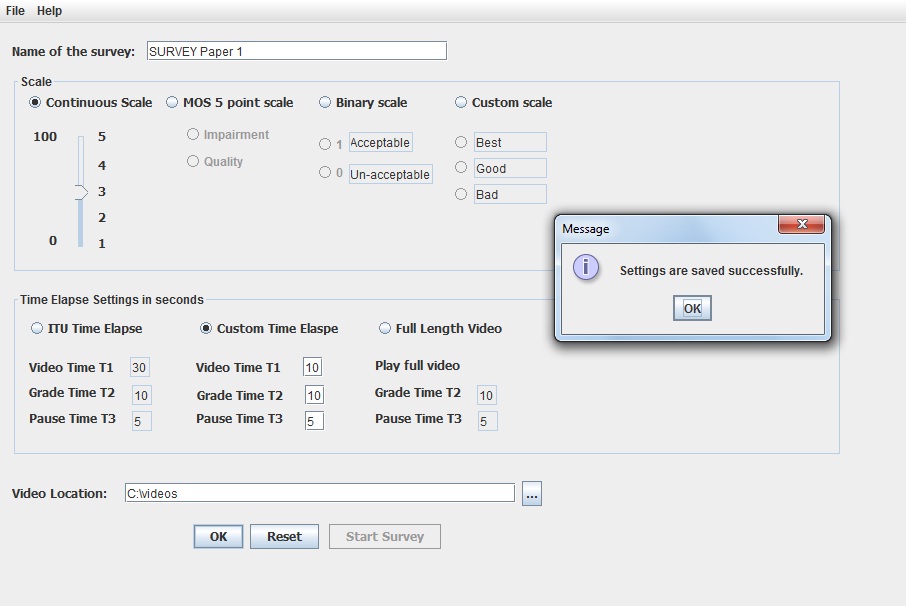}
\caption{Test Configuration of Subjective analysis Tool}
\end{figure}
\begin{figure}[ht]
\centering
\includegraphics[width=0.4\textwidth]{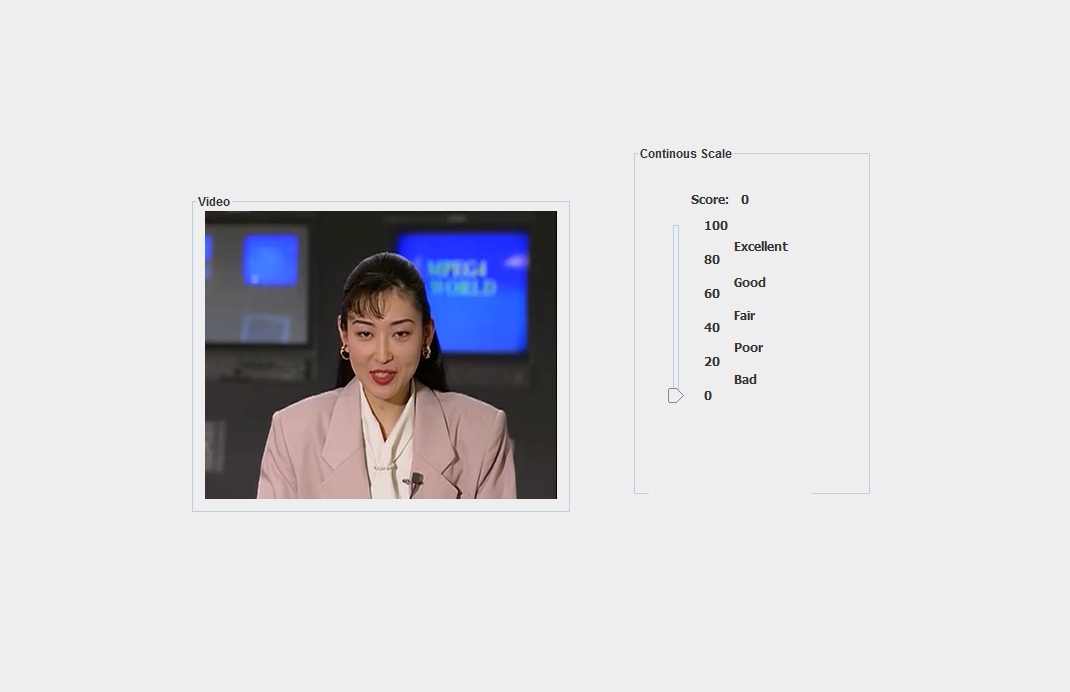}
\caption{Grading Scale of Subjective analysis Tool}
\end{figure}

Non-expert subjects who were invited to participate in the tests were mainly international students in different master programme offered at the university and some staff members also took part in the grading campaign. The viewers were introduced to the tests by dictating a common text saying that they are supposed to grade a set of videos on visual quality basis. To ensure no viewer fatigue, the test sessions were kept around half an hour length. In order to obtain reliable results out of raw subjective scores, a two step filtering method was employed to refine the results. The first step was to detect and discard the observers who exhibited large change of votes compared to the average scores. The second step involved the screening of inconsistent observers without any thought of systematic change. The algorithmic details of these steps are reported  After performing the refining process, the outliers were removed, and we were left with 18 subjects. Mean opinion score (MOS) was calculated from the scores of these subjects for each test condition

\section{2 step Refining Alogirthmn based on H.264 Standards}
We implemnted this alogirthmn to overcome inconsistency within RAW Subjective scores for 27 subjects and even though subjective scores are true values jugded by human subjects after further investigation we understood that mean opnion Scores are considered as independent not dependant variables. We developed a refining algorithm based on H.264 standards while considering ITU recommendations to overcome inconsistency within predicted scores by validating. To Confirm that obtained scores for each time window of test configuration is normal distribution or not, a test was conducted and for achieving first step of refining raw scores, mean, standard deviation and the coefficient for each of all the time windows of each test configuration was computed. This process helps in rejecting observers based on scores significantly far-off from average scores. This step detects and discards the observers based on consistency of votes given and similarly, the distribution of scores is normal or not is confirmed by the means of test. To achieve final step of refining raw scores, mean standard deviation and the coefficient for each of the time windows of each test configuration are calculated. 

\section{ Evaluation of proposed Model}
To obtain the 95\% confidence interval between subjective scores and total number of subjects we need to find upper and lower bounds based on sampling distribution i.e, and now it is possible with normal distribution because our proposed algorithm discared inconsistency with raw scores. So we considered distribution of test configuration for two scenarios i.e, before and after Refining the scores and our model is validted by comparing 95\% confidence for two scenarios as illustrated and following mathematical expression for sampling distribution.
\begin{equation}
95\%CI=tanh(arctanh(r)\pm Z_{\frac{\alpha}{2}}*\frac{1}{\sqrt{n-3}})
\end{equation}
where $Z_{\frac{\alpha}{2}}=1.96$ 
\\ standard deviations, n=number video sequences.

\begin{figure}[h]
\centering
\includegraphics[width=0.3915\textwidth]{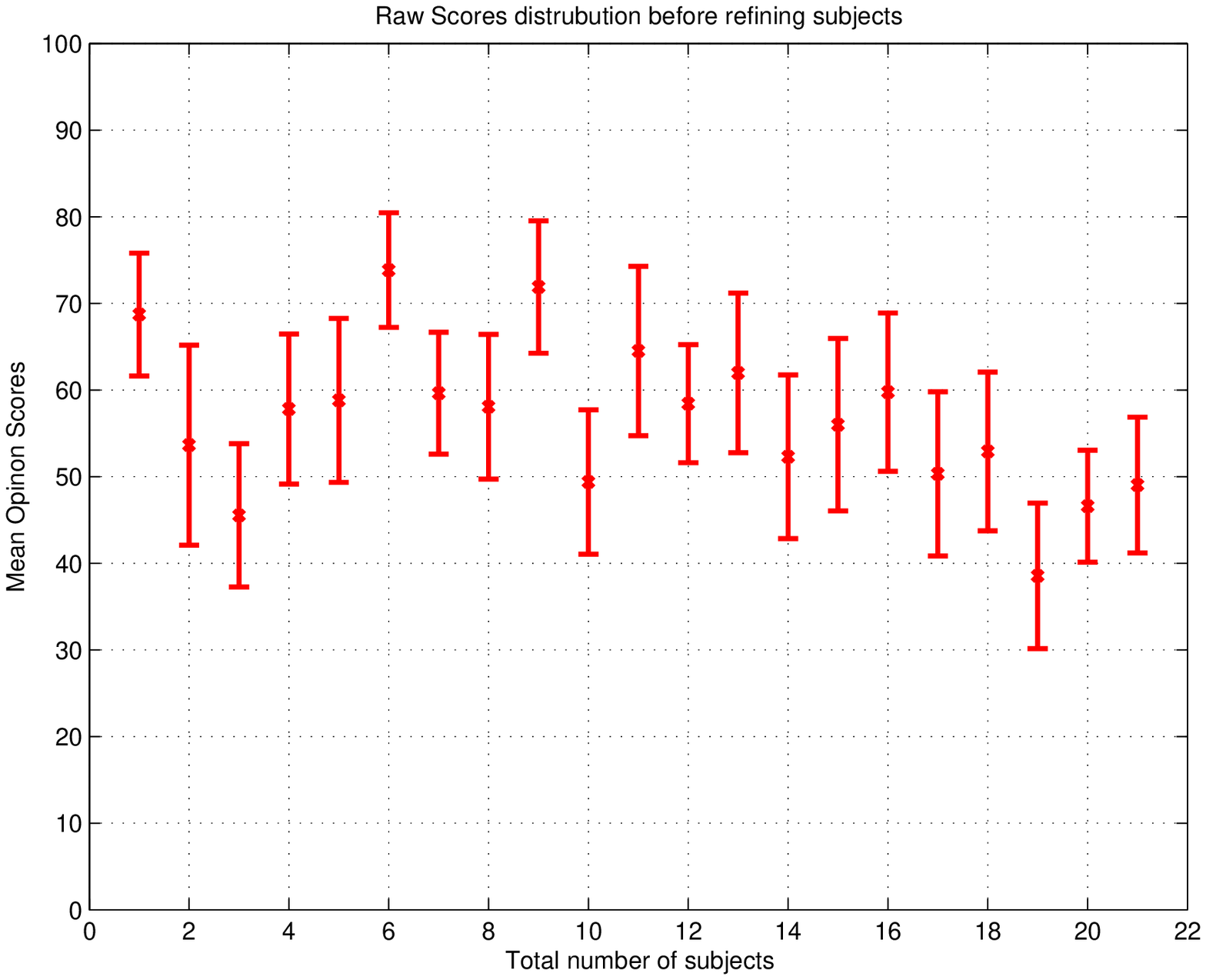}
\centering
\includegraphics[width=0.3915\textwidth]{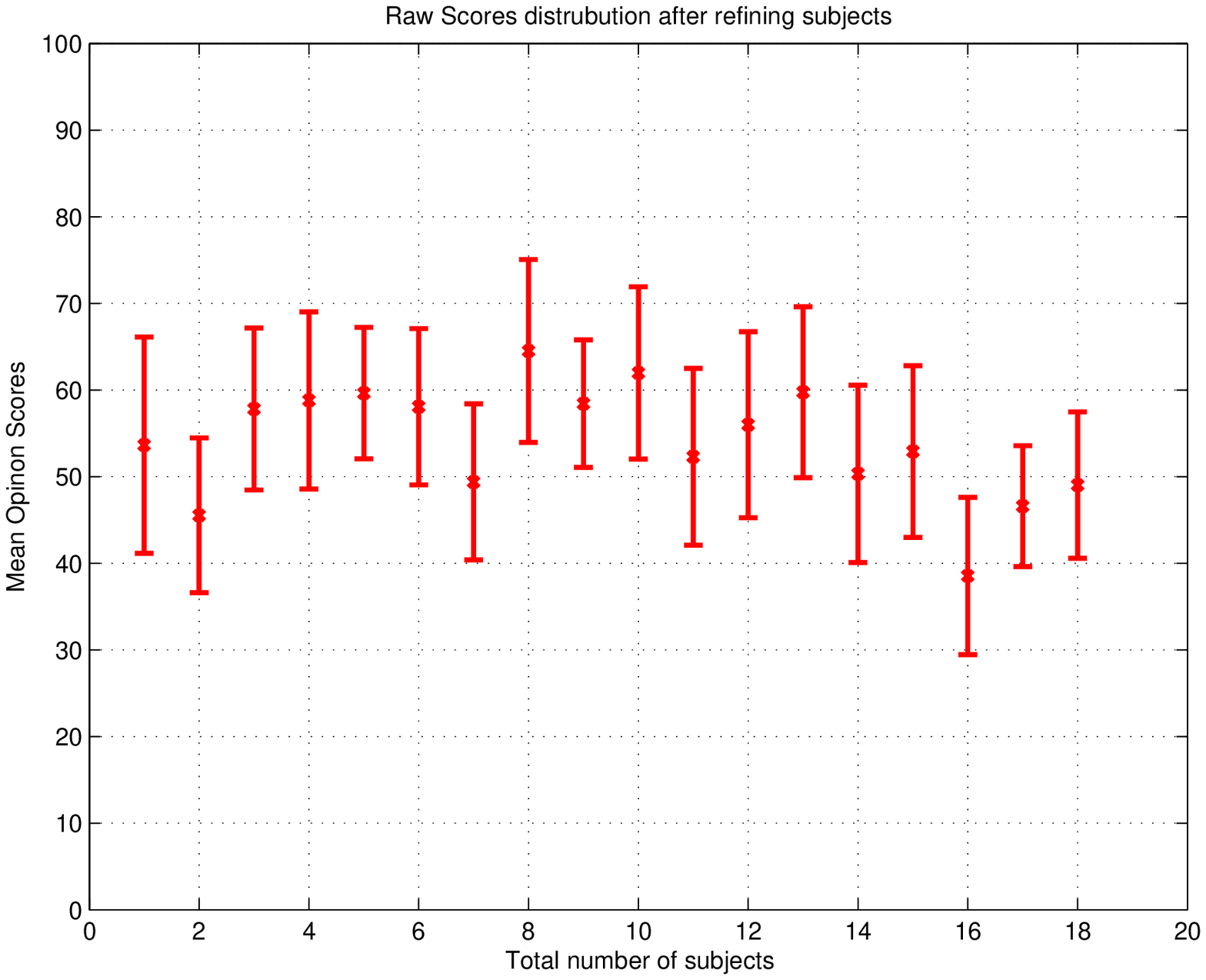}
\caption{95 percent confidence interval of scores before and after refining}
\end{figure}
\section{Validation of coding estimation based on Blahut Algorithm}
 Due to Low Channel Capacity towards inefficient Computation within channel capacity resulted in poor video quality not because of encoding and it was traced out because of subjective quality analysis based on 2 step refining method. As it clearly visible in plotting of before and after refining exploits that data consistency.
 \begin{figure}[!t]
\centering
\includegraphics[width=0.4\textwidth]{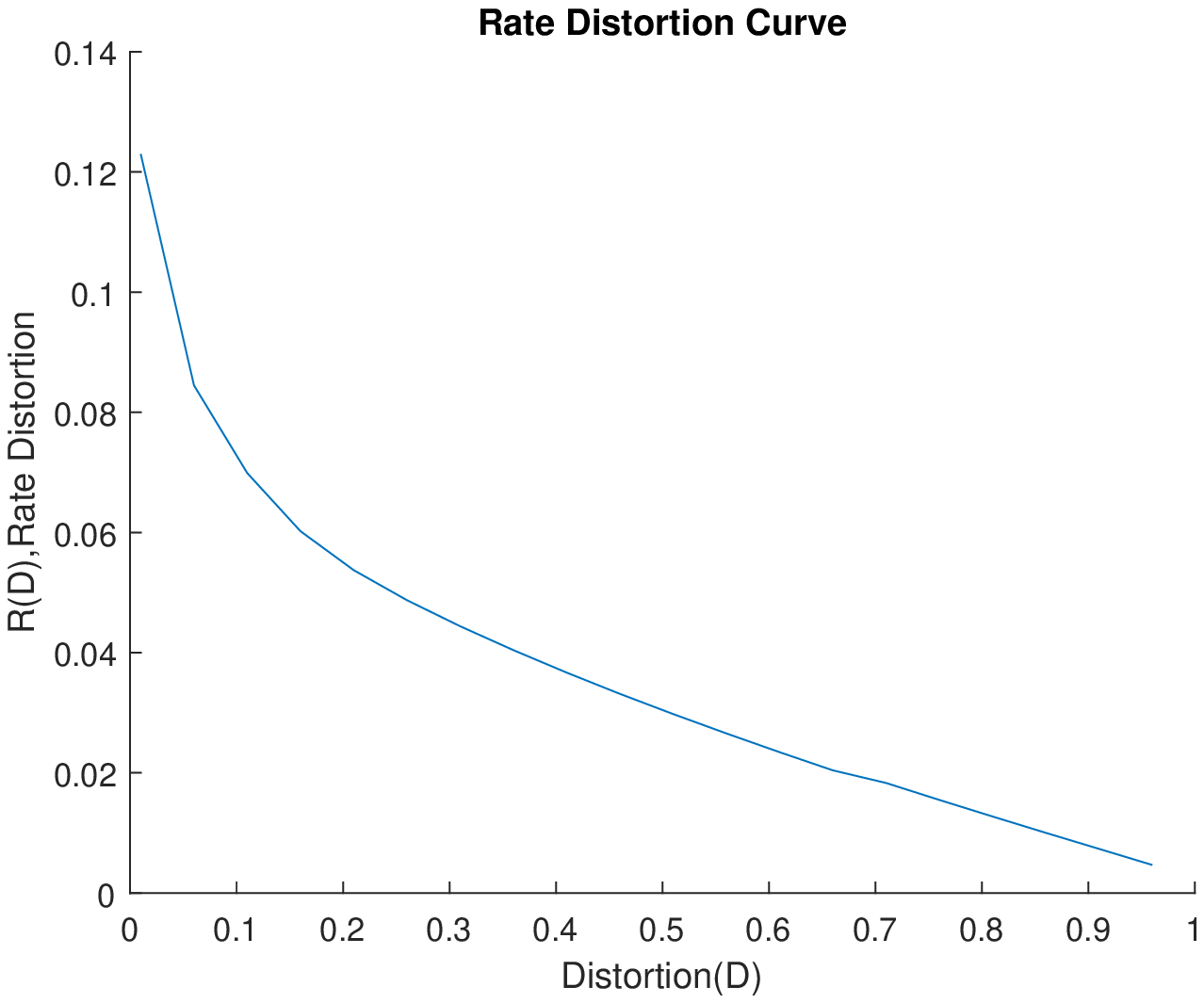}
\includegraphics[width=0.4\textwidth]{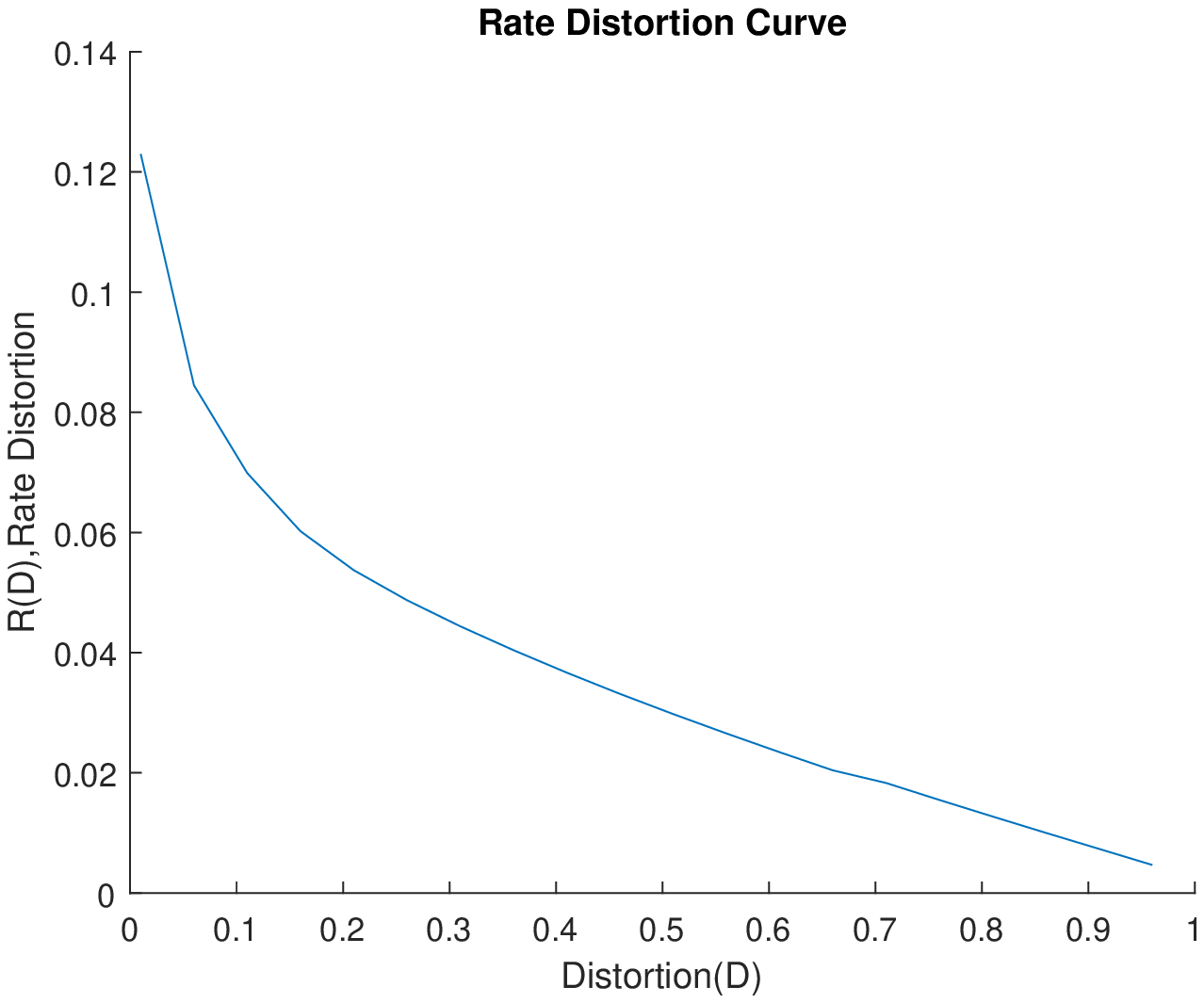}
\caption{Evaluation of rate distortion control for Original or reference video and Reconstructed video}
\end{figure}

\section{Conclusion}
We concluded that after achieving consistency within subjective scores, our investigation stated that human visualization characteristics are considered as true values because of relevant consistency after refining subjective scores, so finally, we understood that it is not a poor coding its due channel capacity which resulted degrading video quality.

\section*{Acknowledgments}
I was confident enough to support and contribute based on my experience and work together with my research group towards achieving good results in the end of my research.


\bibliographystyle{IEEEtrans}
\bibliography{references}
\end{document}